\documentclass[12pt]{article}
\usepackage{epsfig}
\makeatletter
\renewcommand{\section}{\@startsection{section}{1}{\z@}%
{-3.5ex \@plus -1ex \@minus -.2ex}%
{2.3ex \@plus.2ex}{\normalsize\bf}}

\makeatother
\begin{document}
\noindent
{\bf EQUILIBRIUM, ISOSCALING AND NUCLEAR ISOTOPE THERMOMETRY 
RELATED TO 1 GEV PROTON INDUCED REACTIONS}
\\[5mm]
{\bf
M.N.Andronenko, L.N.Andronenko, W.Neubert$^\dagger$ and D.M.Seliverstov}
\\[2mm]
{\it $^\dagger$Institut f\"ur Kern- und Hadronenphysik, 
     FZ Rossendorf, 01314 Dresden, Germany}

\section*{Introduction}
\hspace*{6mm}
Experimental data, related to the decay modes of systems produced in
1 GeV proton-nucleus interactions, have been analysed by methods probed in
nucleus-nucleus collisions. The actual questions are addressed to the next topics:
attainment of equilibrium in fragmenting systems \cite{123}, 
isoscaling \cite{tsang} and nuclear isotope thermometry \cite{albergo,poch}. 
With this intention we have investigated:

$\bullet$ reactions with cumulative production of Light Charged Particles (LCP)

$\bullet$ emission of Intermediate Mass Fragments (IMF) in fragmentation processes

$\bullet$ production of residual nuclei in spallation reactions

The present paper is a review of recent publications \cite{rab,inpc,epja}
based on experimental results from \cite{amn,volnin,aln,batist}.
Especially, the earlier and precise data 
\cite{volnin} are suited for a comprehensive analysis. 
Fragment production in proton collisions with light target nuclei \cite{aln}
allows to test the above mentioned questions including small nuclear systems. 

\section*{Review of experimental data}
\hspace*{6mm}
The data taken into consideration were obtained in several experimental projects 
performed at the external proton beam of the PNPI synchrocyclotron in Gatchina.

$\bullet$ 
One experiment carried out with a lens spectrometer combined with TOF measurements
was aimed to the production of LCPs
detected at backward angles $\Theta_{lab}$=109$^\circ$ and 156$^\circ$ \cite{amn}: 
\begin{center}
$p$(1GeV)+($^6$Li,$^7$Li,$^9$Be,C,Al,$^{58}$Ni,Ag,Pb)$\rightarrow$ LCP($^1$H,$^2$H,$^3$H)+X.\\
\end{center}

$\bullet$ 
The second data set \cite{volnin,aln} which was analysed involves isotopically separated IMF's.
The measurements were performed with a setup consisting of a 
magnetic lens spectrometer equiped with $\Delta$E-E telescopes 
at $\Theta_{lab}$=60$^\circ$ and 120$^\circ$ with respect to the beam axis.
Fragment production in light targets was investigated with two 
TOF-E spectrometers including PPAC's and  Bragg ionization chambers
installed at $\Theta_{lab}$=30$^\circ$ and 126$^\circ$. 
The  following data were obtained in these experiments: 

\hspace{1cm} - differential cross sections at forward and backward angles:
\begin{center}
$p$(1GeV)+($^9$Be,C,$^{58}$Ni,Ag,Au,$^{238}$U)$\rightarrow$IMF($Z\geq$2)+X.\\
\end{center}

\hspace{1cm} - differential cross sections at $\Theta_{lab}$=60$^\circ$:
\begin{center}
$p$(1GeV)+($^{48}$Ti,$^{58}$Ni,$^{64}$Ni,$^{112}$Sn,$^{124}$Sn)$\rightarrow$IMF($Z\geq$2)+X.
\end{center}

$\bullet$ 
The yields of spallation products at E$_p$=1 GeV incident energy were 
measured with a high-resolution gamma spectrometer \cite{batist}. 
These copious data allowed us to extend the analysis 
toward heavier isotopes up to residual nuclei with masses 
close to the target mass A$_T$ considering the reactions: 
\begin{center}
$p$(1GeV)+($^{51}$V,$^{54,56}$Fe,$^{58,60,62,64}$Ni,
$^{70,76}$Ge,$^{185}$Rb,$^{109}$Ag,$^{133}$Cs)$\rightarrow$ Residue+X.\\
\end{center} 

\section*{Probe the equilibration in fragmentation and spallation reactions}
\hspace*{6mm} 
One important question for the understanding of the nuclear disintegration
is the search for signals of equilibration of the emitting source. One access
is related to the isotopic yield ratios of the reaction products \cite{rab}.
In the grand canonical approach, the yield ratio {\bf R} of 
{\em two} different isotopes 
($Z,N_1$) and ($Z,N_2$) emitted from {\em one} source is given by:     
\begin{equation}
{\bf R} =\left(\frac{N_2+Z}{N_1+Z}\right)^{3/2}\
\cdot \exp\left(-\frac{\varepsilon_2-\varepsilon_1}{T}\right)  \nonumber\\
\cdot \exp\left(\frac{\mu_n(N_2-N_1)}{T}\right) 
\end{equation}
where $T$ is the equilibrium temperature, $\varepsilon$ is the mass excess 
and $\mu_n$ is the chemical potential. 
The chemical potential of the neutrons is assumed to be
a linear function of the neutron-to-proton ratio of the combined system
of the target($Z_t$,$N_t$) + projectile($Z_p$,$N_p$). 
Then for proton-induced reactions, the yield ratio of two isotopes with
$N_1$ and $N_2$=$N_1$+$\Delta N$ produced at the same temperature $T$ can be
expressed by
\begin{equation}
{\bf R} \propto \exp(c_o \cdot \Delta N \cdot N_t / (Z_t +1) )
\end{equation}
According to (2), in case of equilibrium 
the ratios lie in a semi-log plot {\bf R} versus $N_t$/($Z_t$ +1)
on a straight line as one can see in Figs.~1 and ~2. This is valid for 
both the fragmentation and the spallation reaction products provided that
$\Delta N$ is kept fixed.
\vspace{-20mm}   
\begin{center}
\begin{tabular}{cc}
\epsfig{file=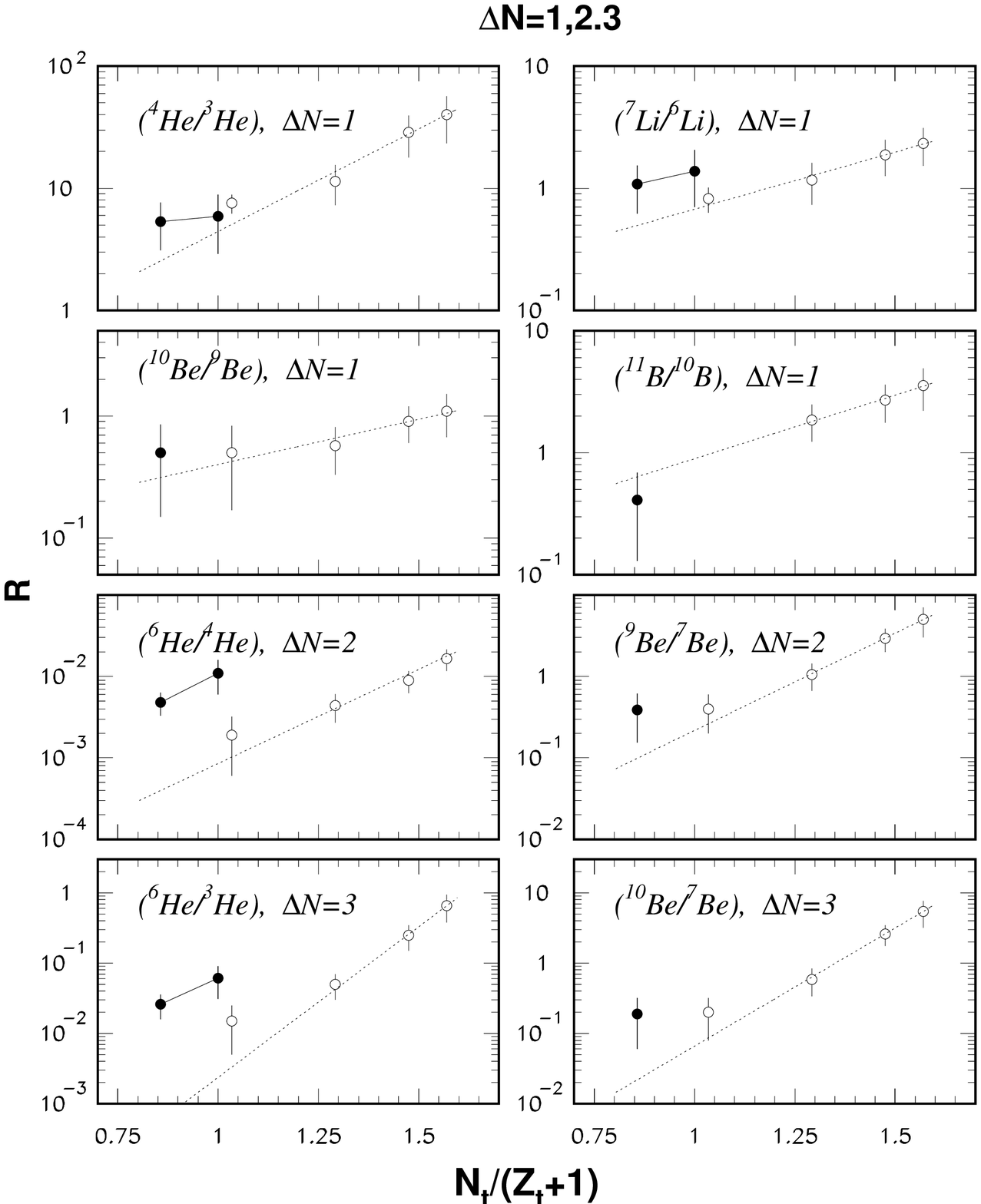,width=8cm,height=8.0cm}
\noindent
\small
&
\epsfig{file=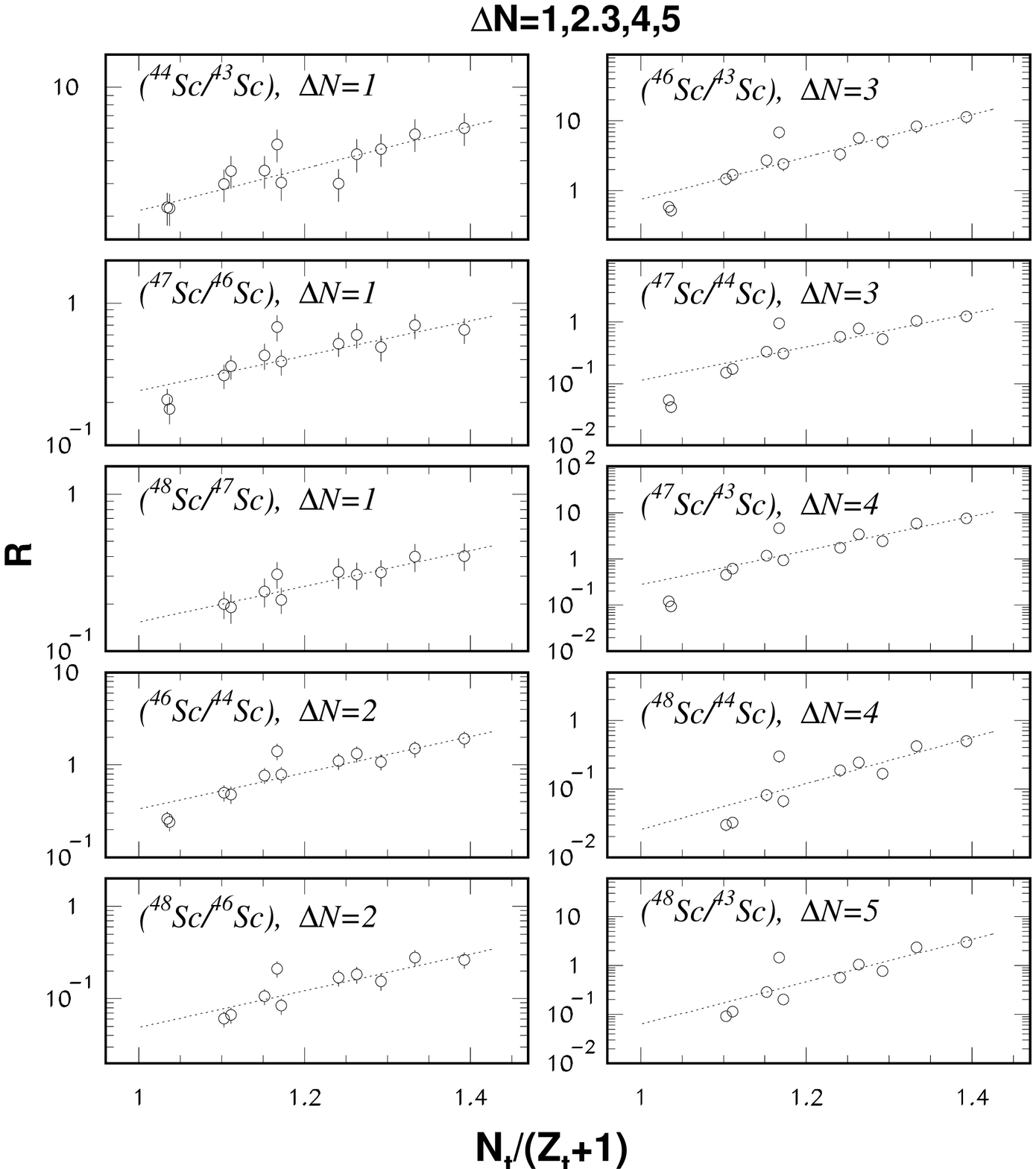,width=9cm,height=9.0cm}
\noindent
\small
\\
\begin{minipage}{8cm}
{\small {\bf Fig.~1.} Isotopic yield ratios in $p$(1 GeV)+$A$ 
fragmentation reactions for target nuclei from $^9$Be to $^{238}$U.
The lines are exponential fits to the experimental points (open circles), 
the ratios for light targets are denoted by black dots}
\end{minipage}
\normalsize
&
\begin{minipage}{8cm}
{\small {\bf Fig.~2.} Yield ratios {\bf R} of $^{43-48}$Sc isotopes 
obtained in spallation reactions induced by \mbox{1 GeV} protons 
with the target nuclei from $^{51}$V to $^{133}$Cs.
The same notation as in Fig.~1 is used}
\end{minipage}
\normalsize
\end{tabular}
\end{center}
\vspace{5mm}   
The slope $K({\Delta N})= c_o \cdot {\Delta N}$ is proportional to the difference  
of the neutron numbers $\Delta N$ \cite{123} and fulfils the condition
\begin{equation}
K({\Delta N})_{\Delta N = 1, 2, 3, 4, ...} = 1 : 2 : 3 : 4 ...
\end{equation}
The confirmation of relation (3) for  
two classes of reactions is demonstrated in Fig.~3.
   The fact that the ratios related to light nuclei 
(see black dots in Fig.~1) don't coincide with the fit lines 
responsible for heavy nuclei may be explained by 
differences in $N/Z$ of the combined system and the 
actual ratio $N/Z$ of the fragmenting system.
   Note that in the case of light disintegrating nuclei the values of $N/Z$
are very sensitive to changes 
of the proton and neutron constituents by one unit.  
The slope of ${\bf R}(N_t / (Z_t +1))$ dependency 
for light targets are similar to one for heavy targets.
\vspace{-0.7cm}
\begin{figure}[h]
\begin{minipage}[b]{8.0cm}
  This observation may be an indication in favour of 
a similar fragment production mechanism in light nuclei,
i.e. there is no reason to refuse the statistical 
treatment completely.

\underline{Summarizing}, we found for fragmentation and deep inelastic reactions
(spallation) at 1\,GeV incident proton energy a consistent behavior which may be
attributed to attained equilibration characterized by an effective temperature.
The exponential slopes of the isotopic yield ratios 
for fixed difference of the neutron numbers $\Delta N$=1,2,3,...
were found to be nearly independent of  
the masses of both the fragments and the emitting system.
\end{minipage}\hfill
\begin{minipage}[b]{8.0cm}
\begin{center}
\epsfig{file=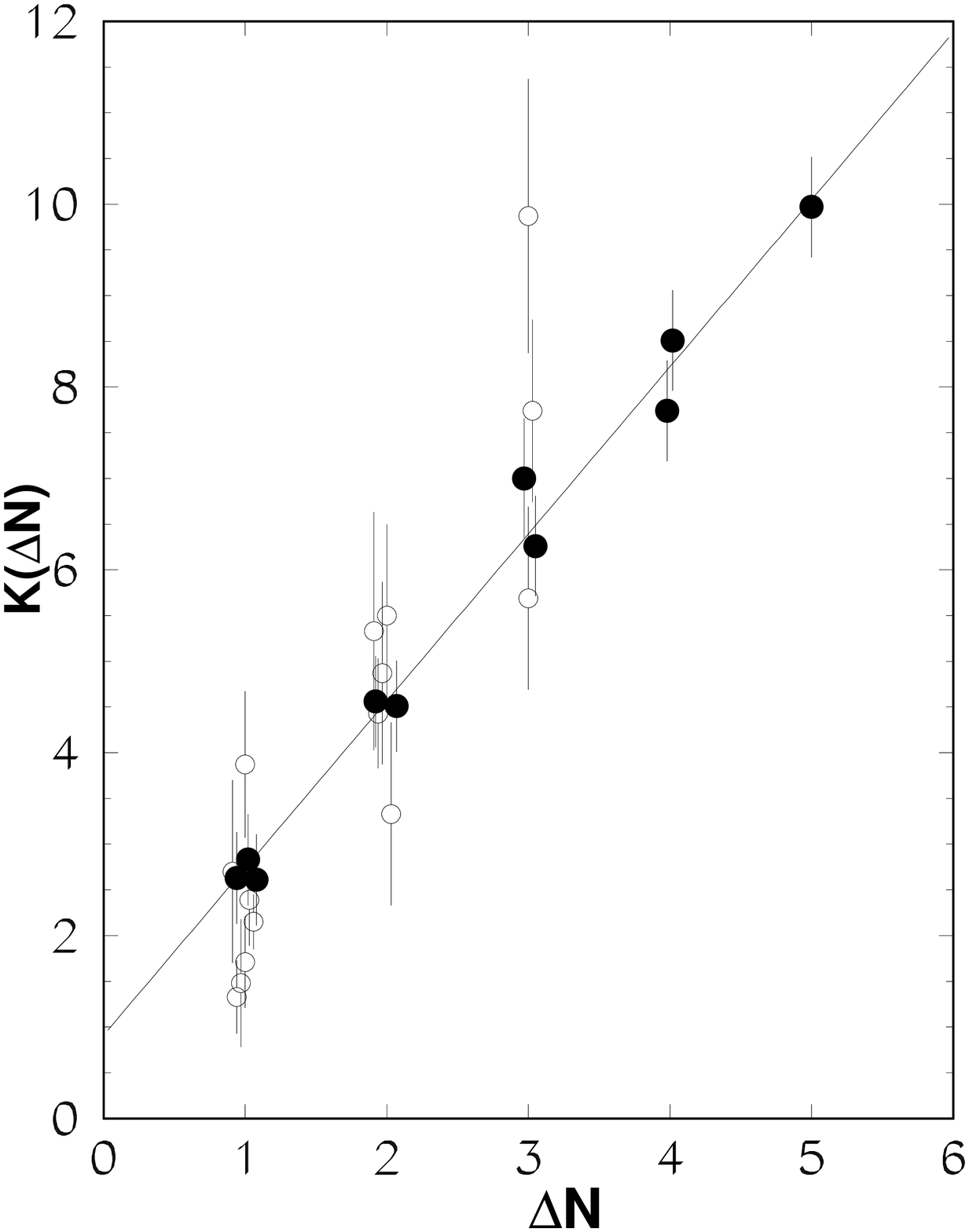,height=5.9cm}\\ 
\end{center}
\vspace{-5mm}
\small {\bf Fig.~3.} 
The slopes $K({\Delta N}$) as function of the difference of
the neutron numbers of the reaction products from fragmentation 
(open points) and spallation (solid points). 
The line is the fit to spallation data
\end{minipage}
\end{figure}

\section*{Isotopic scaling and its dependence on the isospin of the emitting source}
\hspace*{6mm}
Supposing that equilibrium is attained, the grand canonical 
expression for the yield ratio of {\em specific} isotopes with $N,Z$ 
from two {\em different} emitting systems 
can be expanded to first order in $N$ and $Z$ \cite{tsang} :
\vspace{-2mm}
\begin{equation}
Y_2/Y_1=C \cdot  \exp(\alpha N + \beta Z) 
\end{equation}
where $\alpha=\Delta \mu_n/T$, $\beta=\Delta \mu_p/T$ and
$\Delta \mu_n$=$\mu_{n2} -\mu_{n1}$ ($\Delta \mu_p$=$\mu_{p2} -\mu_{p1}$)
are the differences of the neutron (proton) chemical potentials for  
the two emitting systems and $T$ is the equilibrium temperature.
In such a case the scaled isotopic ratios
\begin{equation}
S(N)=Y_2/Y_1  \cdot \exp(-\beta Z)
\end{equation}
lie along a straight line on a semi-log plot $S(N)$ as function of $N$. 
This three-parametric isotopic scaling (called isoscaling) has been demonstrated
in the restricted range 0 $\leq N \leq$ 11 for multifragmentation, 
strongly damped binary collisions and evaporation \cite{tsang}. 

We have established the validity of such scaling behavior 
also for proton induced reactions in the GeV energy region ~\cite{inpc}.  
The isotopic yield ratios for fragmentation products (2$\leq Z \leq$ 5) and 
spallation residues \mbox{(11$\leq Z \leq$ 28)} obtained from numerous targets, 
irradiated with 1 GeV and 12 GeV protons, were analysed. 
Many combinations of systems for both reaction types were considered. 
Figure 4 demonstrates the scaling behavior in the range of neutron numbers 
20 $\leq N \leq$ 33 for proton induced spallation for one chosen combination 
of two emitting systems. A complete picture of combinations of yield ratios 
available for spallation and fragmentation products is shown in Figs.~5 and ~6.  
Obviously, the parameters $\alpha$ and $\beta$ are different 
for various system combinations $Y_2/Y_1$.
\vspace{-5mm}
\begin{center}
\begin{tabular}{cc}
\epsfig{file=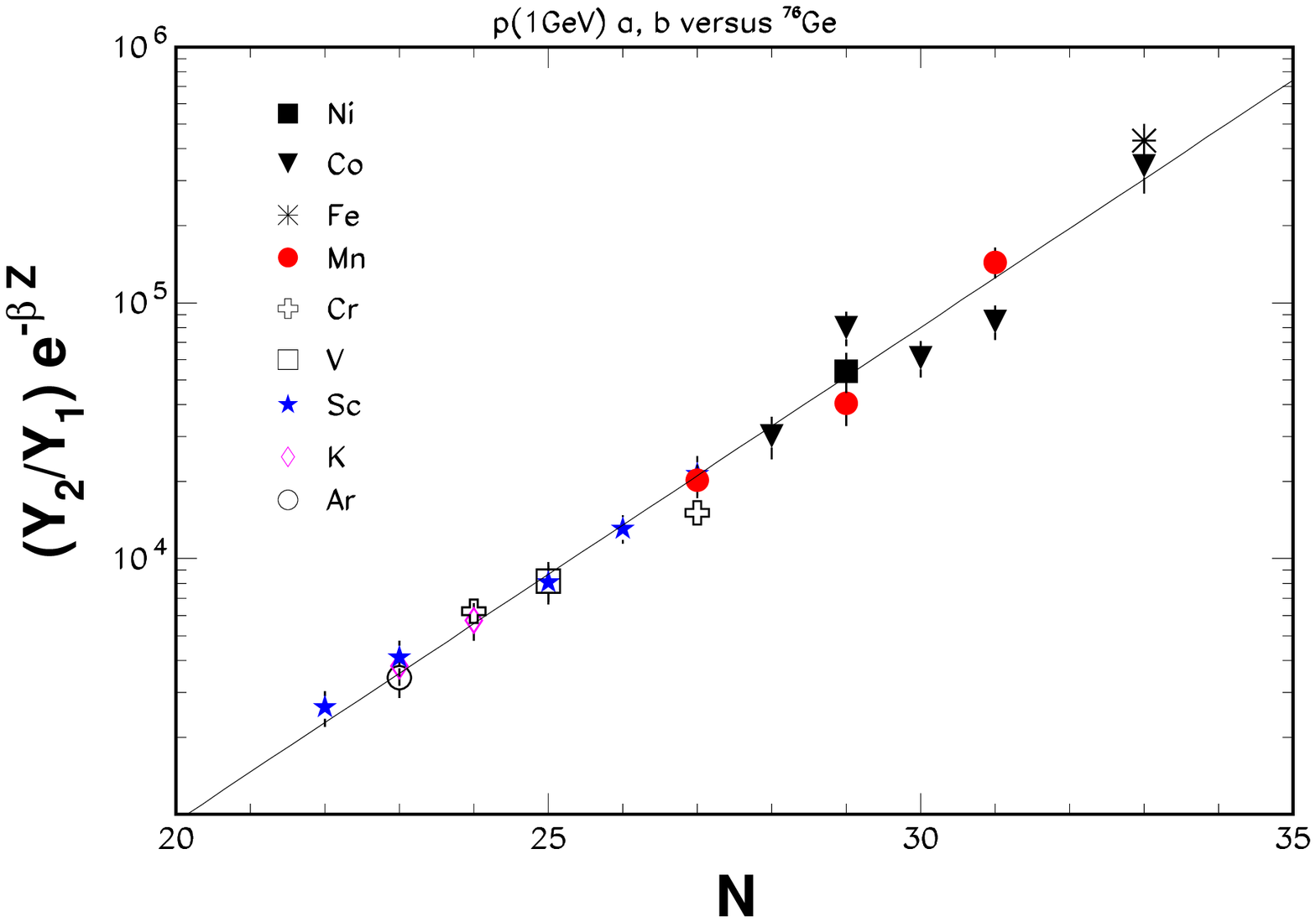,width=7.7cm} 
\noindent
\small
&
\epsfig{file=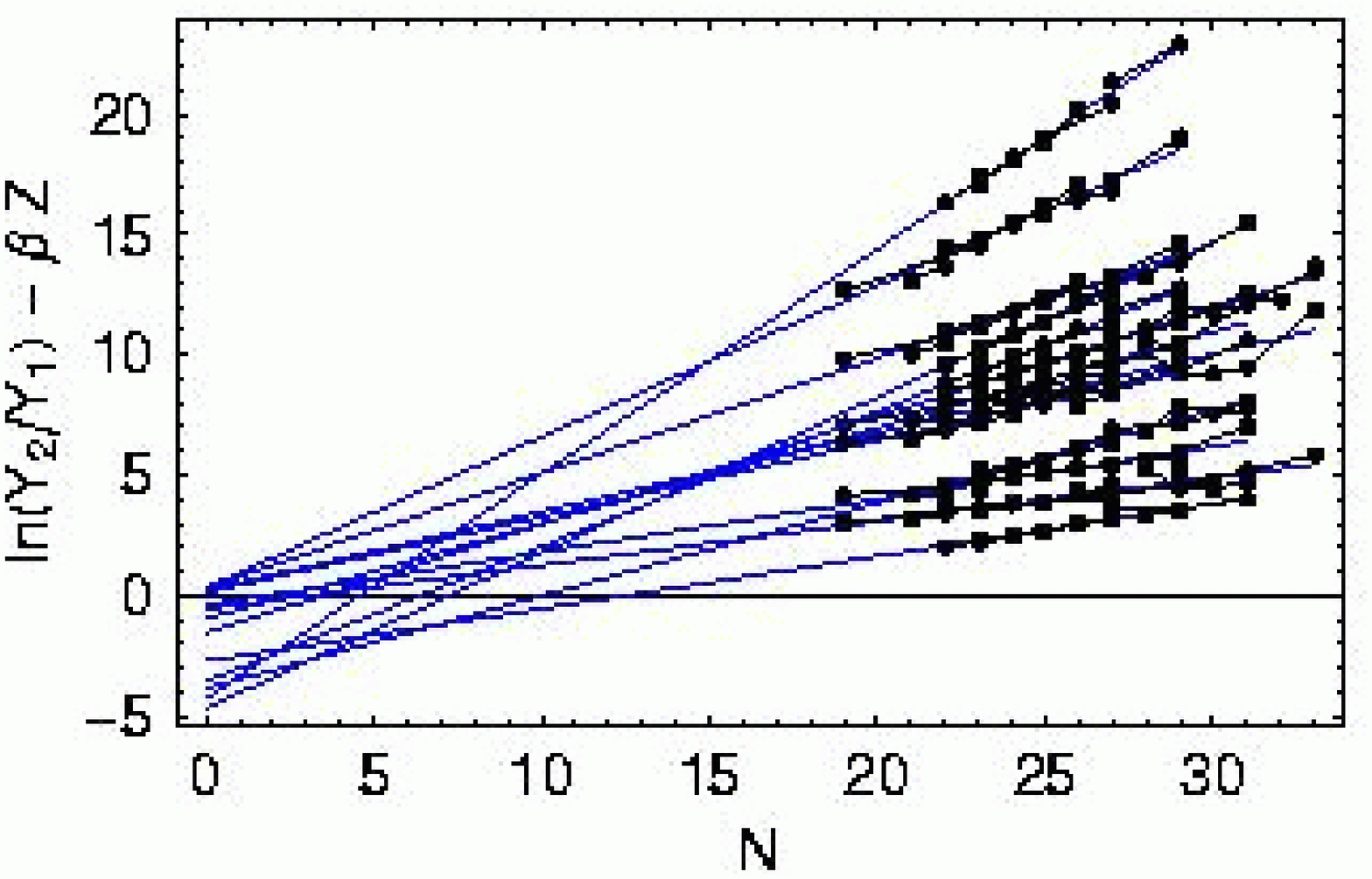,width=8cm} 
\noindent
\small
\\
\begin{minipage}{8cm}
{\small {\bf Fig.~4.} 
Scaled isotopic ratios $S(N)$ as function of the neutron number
of the spallation residues. The plot includes isotopes of elements  
which are produced in $p$(1GeV) + $^{70,76}$Ge interactions \cite{batist} }
\end{minipage}
\normalsize
&
\begin{minipage}{7.5cm}
{\small {\bf Fig.~5.} Spallation at 1 GeV proton energy. The 
yields of reaction products of proton collisions with the 
following target nuclei are involved: 
$^{58,60,62,64}$Ni, $^{70,76}$Ge \cite{batist} } 
\end{minipage}
\normalsize
\end{tabular}
\end{center}
\vspace{-2mm}
\begin{figure}[h]
\begin{minipage}[b]{8.5cm}
The coefficients $\alpha$ and $\beta$ were found to be 
dependent on the isoscaling factor
\begin{equation}
\Delta \xi ={\mathcal N}_{2}/({\mathcal Z}_{2}+1)
 -{\mathcal N}_{1}/({\mathcal Z}_{1}+1)
\end{equation}
Here, ${\cal N}_{1}$, ${\cal Z}_{1}$ and ${\cal N}_{2}$, ${\cal Z}_{2}$ 
are the neutron and proton numbers of the source '1' and source '2', respectively.
For simplicity, we assume that the nucleonic composition of the
emitting source is nearly the same as in the system p+target.  

On this condition we found a linear dependence 
of $\alpha $ and $\beta$ on $\Delta \xi$:
\begin{equation}
\alpha = \alpha ^{'} \cdot \Delta \xi \, \hspace*{3mm}and \, \hspace*{3mm}
\beta = \beta ^{'} \cdot \Delta \xi
\end{equation}
\end{minipage}\hfill
\begin{minipage}[b]{7.5cm}
\begin{center}
\epsfig{file=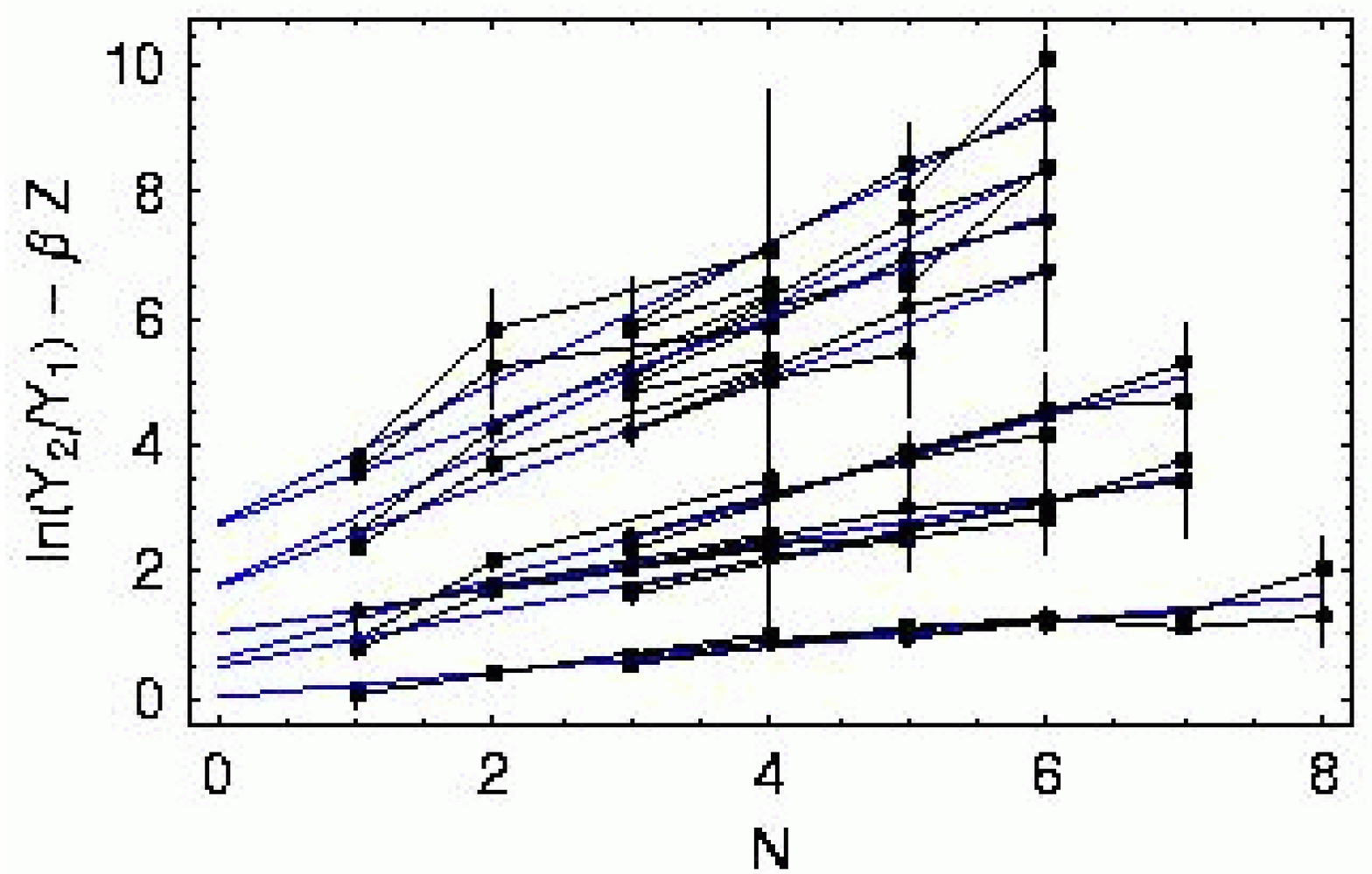,width=7.2cm}\\
\end{center}
\vspace{-5mm}
\small {\bf Fig.~6.} Fragmentation at 1 GeV proton energy. 
The yield ratios of IMFs for the target nuclei from uranium 
to carbon are used: \cite{volnin,aln} 
\end{minipage}
\end{figure}
Figures ~7 and ~8 show the parameters $\alpha $ and $\beta$  
versus $\Delta \xi$ together with the linear fits
from which $\alpha ^{'}$ and $\beta ^{'}$ were obtained.
\vspace{-0.5cm}
\begin{center}
\begin{tabular}{cc}
\epsfig{file=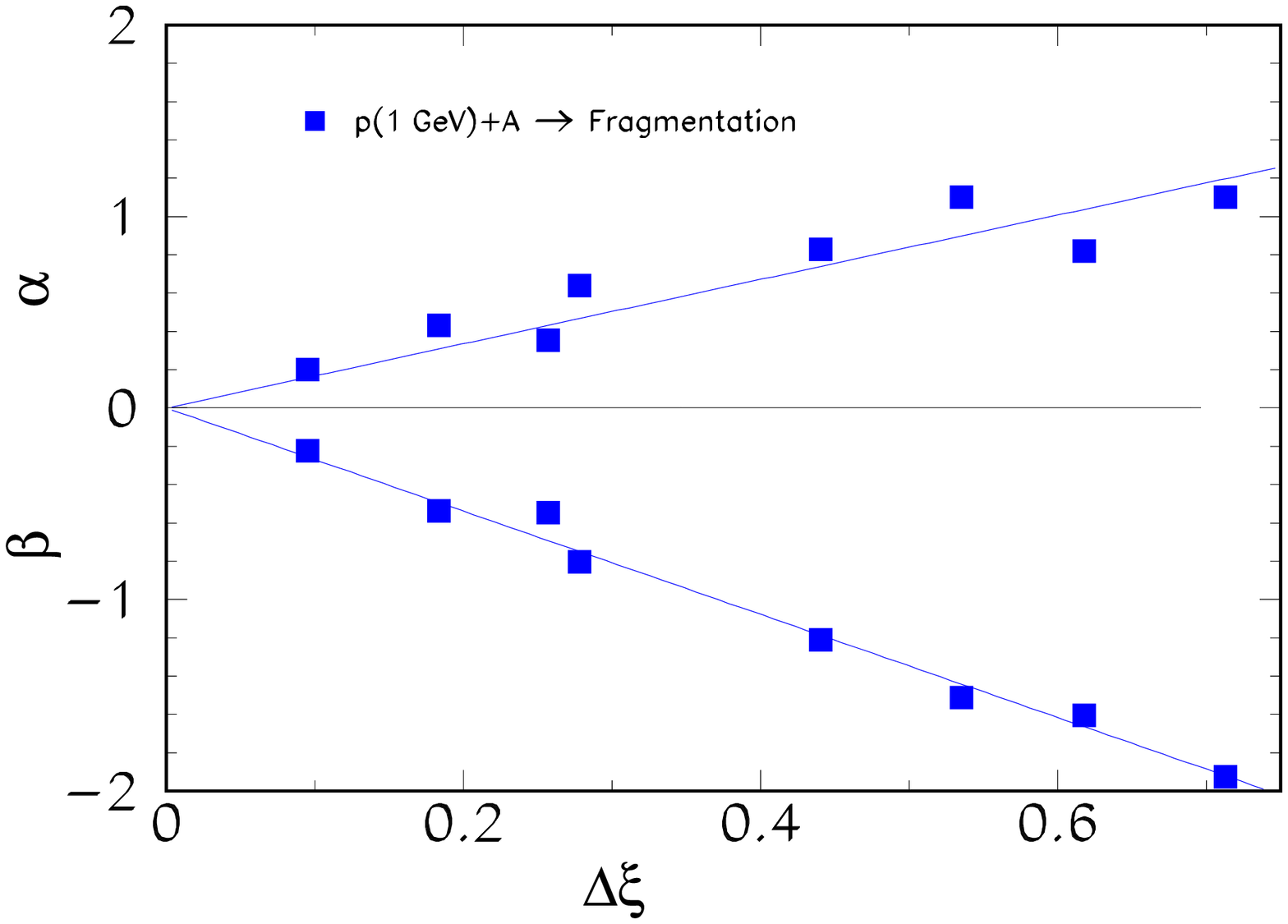,width=7.8cm} 
\noindent
\small
&
\epsfig{file=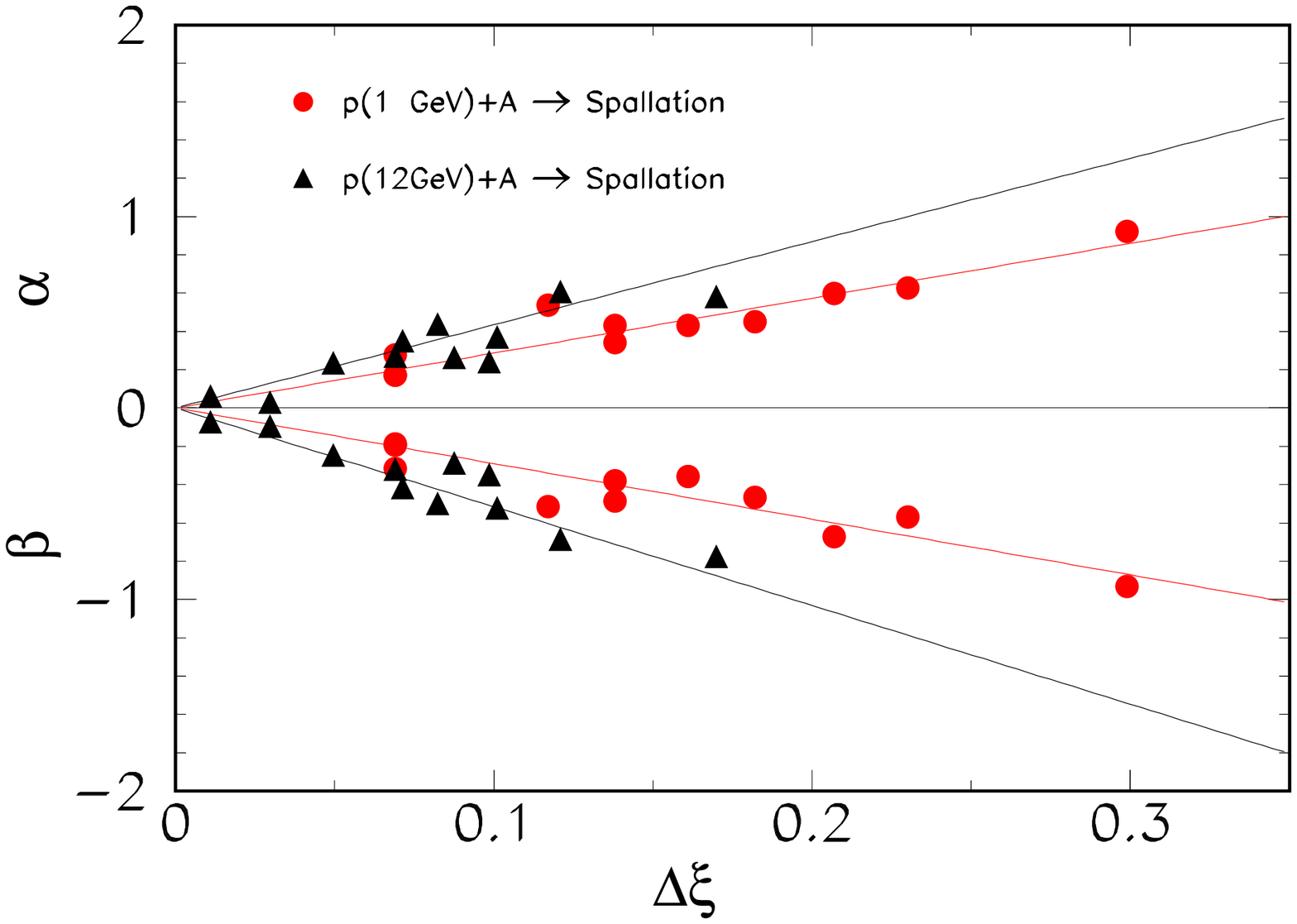,width=7.8cm} 
\noindent   
\small
\\
\hspace{-5mm}
\begin{minipage}{9.0cm}
{\small {\bf Fig.~7.}  
The scaling parameters as function of the 
factor $\Delta \xi$. Fragmentation 
data from \cite{volnin,aln} }
\end{minipage}
\normalsize
&
\begin{minipage}{6.5cm}
{\small {\bf Fig.~8.} The same as in Fig.~7 for spallation data \cite{batist,nogu} }
\end{minipage}
\normalsize
\end{tabular}
\end{center}
Using relations (6) and (7) the equation (4) can be corrected for the 
isoscaling factor $\Delta \xi$ of the two systems under consideration:
\begin{equation}
Y_2 /Y_1=C^{'} \cdot \exp((\alpha ^{'} N + \beta ^{'} Z) \cdot \Delta \xi )
\end{equation}
\vspace{-10mm}
\begin{figure}[h]
\begin{minipage}[b]{8.5cm}
In the following we use modified scaled isotopic ratios: 
\begin{equation} 
S^{'}(N)=(\Delta \xi)^{-1} \cdot (\ln(Y_2/Y_1) -\beta Z)
\end{equation} 
The corresponding results are illustrated in Fig.~9. In this case, the parameters
obtained by fits of the relations (7) are taken into account.
A pronounced scaling behavior becomes evident for fragmentation and spallation
products. Figure 9 shows that all sets of linear dependences presented in Figs.~5 and ~6 
can be transformed into corresponding uniform dependences with similar slopes.
On the other side, the data related to the yields of spallation products obtained 
with 12 GeV protons form a separate line displayed in the upper part of Fig.~9.
\end{minipage}\hfill
\begin{minipage}[b]{7.5cm}
\begin{center}
\epsfig{file=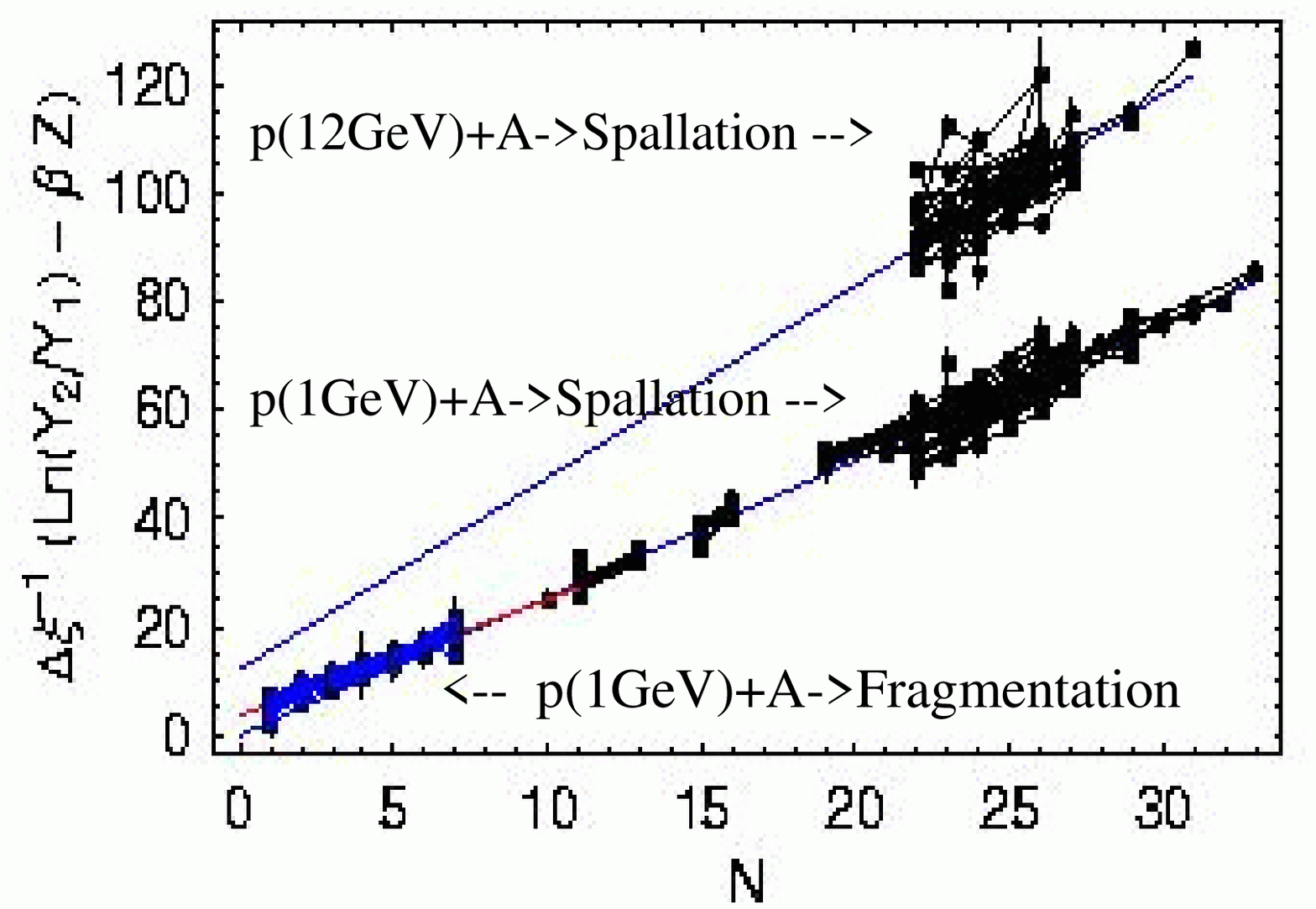,height=6cm} \\
\end{center} 
\vspace*{-5mm}
\small {\bf Fig.~9.} 
Factorization of the isotopic ratios of fragmentation and spallation 
products from $p$ + $A$ collisions corrected for the factors $\Delta \xi$ 
according to relation (9). Lower data points: \cite{volnin,aln,batist},
upper points: \cite{nogu}
\end{minipage}  
\end{figure}

\underline{Summarizing} the results of this section one can conclude:

$\bullet$ Scaling relationships of isotopic distributions have been observed 
          for both reaction mechanisms, i.e. proton induced fragmentation and spallation

$\bullet$ The influence of the isospin of the emitting systems 
          on the isoscaling parameters $\alpha$ and $\beta$ 
          has been established

$\bullet$ The scaling parameters $\alpha ^{'}$ and $\beta ^{'}$ vary with 
          the energy of the primary proton beam

\section*{An analysis of proton induced fragmentation using double isotopic yield ratios}
\hspace*{6mm}
The double yield ratio {\em R = R$_1$/R$_2$} 
of isotopes is related to the nuclear temperature $T_{app}$ 
as shown by Albergo and coworkers \cite{albergo}:
\vspace{-2mm}
\begin{center}
$T_{app} = B\,/\,\ln(a \cdot R)$
\end{center}
where the single isotope ratios are defined by the corresponding yields Y 
\begin{center}
$R_1 = Y(A_i,Z_i)\,/\,Y(A_i+\Delta A,Z_i+\Delta Z)$\\
$R_2 = Y(A_j,Z_j)\,/\,Y(A_j+\Delta A,Z_j+\Delta Z)$
\end{center}
Each combination of four isotopes terms an 'isotope thermometer'.
The quantity $B$ is the difference of the binding energies related to the
above given ratios. The parameter $a$ includes the 
spin degeneration factor and mass numbers of the considered isotopes.

This approach became well-known with the pioneering studies in ref.\cite{poch}
aimed to prove the nuclear phase transition in nuclear matter. 
We applied this kind of analysis to data available from inclusive measurements 
at 1 GeV proton interactions with various target nuclei 
to study the mass dependence of $T_{app}$ \cite{epja}.
This method provides the most reliable temperature estimations
(so called 'isotopic temperature')
as far as the excitation energy of the emitting system
is restricted to \mbox{ $\simeq 3\,-\,5\,A\cdot$MeV}.

Concerning LCP production at 1 GeV proton energy, we assumed
that cumulative particles originate from some 
statistical-like process.
Figure 10 shows the results obtained with the LCP thermometer 
($^2$H/$^3$H)/($^1$H/$^2$H) whereby the yields of the hydrogen
isotopes were measured at two angles.
The temperature $T_{app}$ is nearly constant at $\Theta_{lab}$=109$^\circ$ within
the range of target masses 6$\leq $A$_T \leq$ 208.
The lower part of Fig.~10 shows temperatures 
which are derived from the differential cross sections
of hydrogen isotopes detected at $\Theta$=156$^\circ$. 
It should be noted that an increase of $T_{app}$ at A$_T \leq$ 10 cannot be excluded.
Corrections for the \mbox{$\Delta$-isobaric} state contributions
to the differential cross section (as shown in ref.\,\cite{amn}) 
amount to $\leq$ 20\%,  
but it does not change the tendency of increasing $T_{app}$.

Additional examples of temperatures obtained from He and IMF yields  
are plotted in Fig.~11 as function of A$_T$. 
This picture confirms the regular behavior of temperatures displayed in Fig.~10.
The observed agreement of the studied thermometers implies that 
possibly each of them is suitable for relative temperature measurements
without the hitherto introduced limitation {\em B} $\geq$ 10 MeV. This condition
was introduced in ref.\cite{tsang}
to minimize fluctuations due to contributions to the isotopic yields from
sequential decays. For a given thermometer the influence of
sequential decays seems to be independent on the origin of the
excited primordial fragments. This behavior is rather surprising since the 
target mass numbers (or the volumes of the fragmenting nuclei, respectively)
change by a factor of about 25. 

\vspace{-10mm}   
\begin{center}
\begin{tabular}{cc}
\epsfig{file=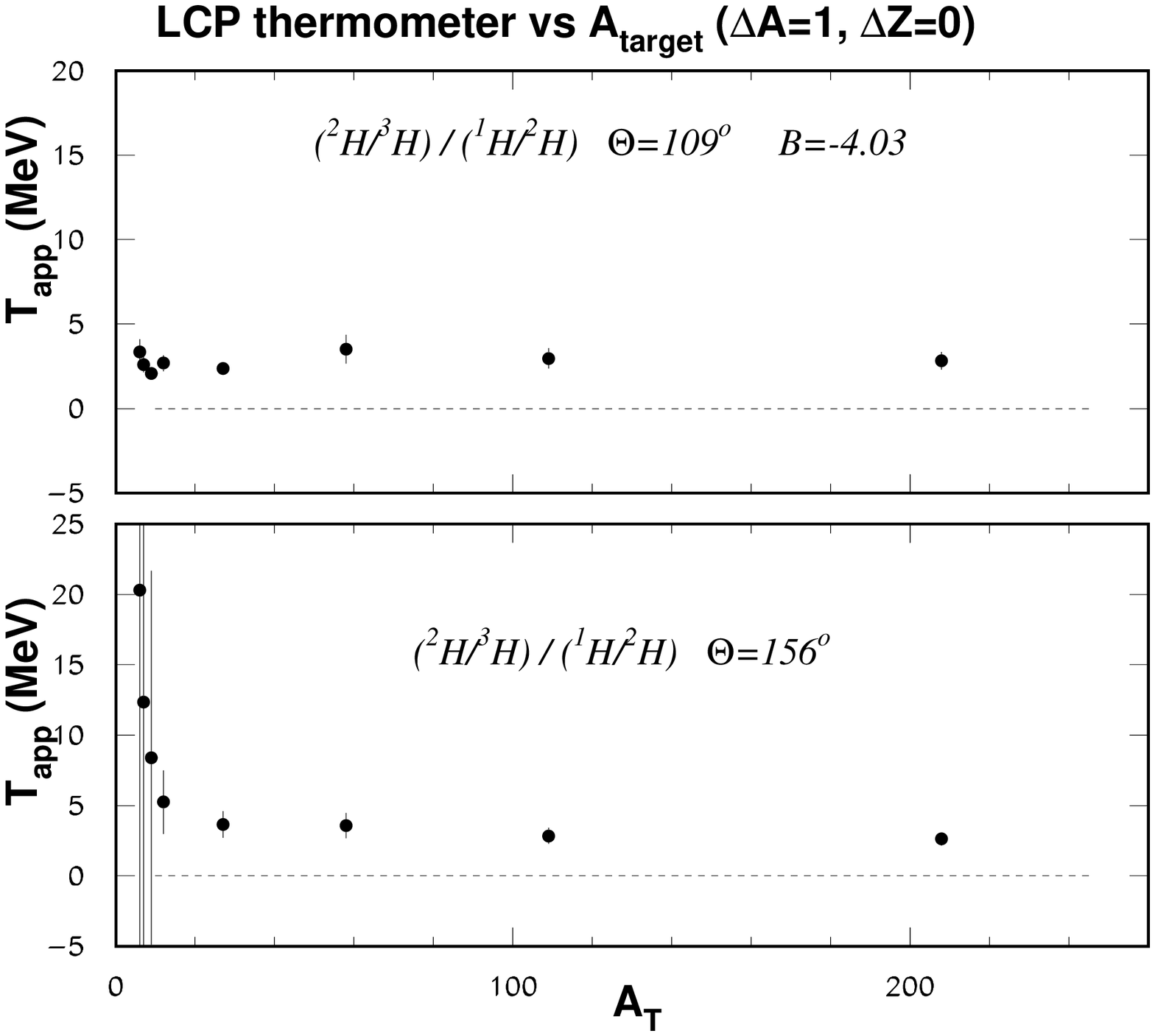,height=8cm}
\noindent
\small
&
\epsfig{file=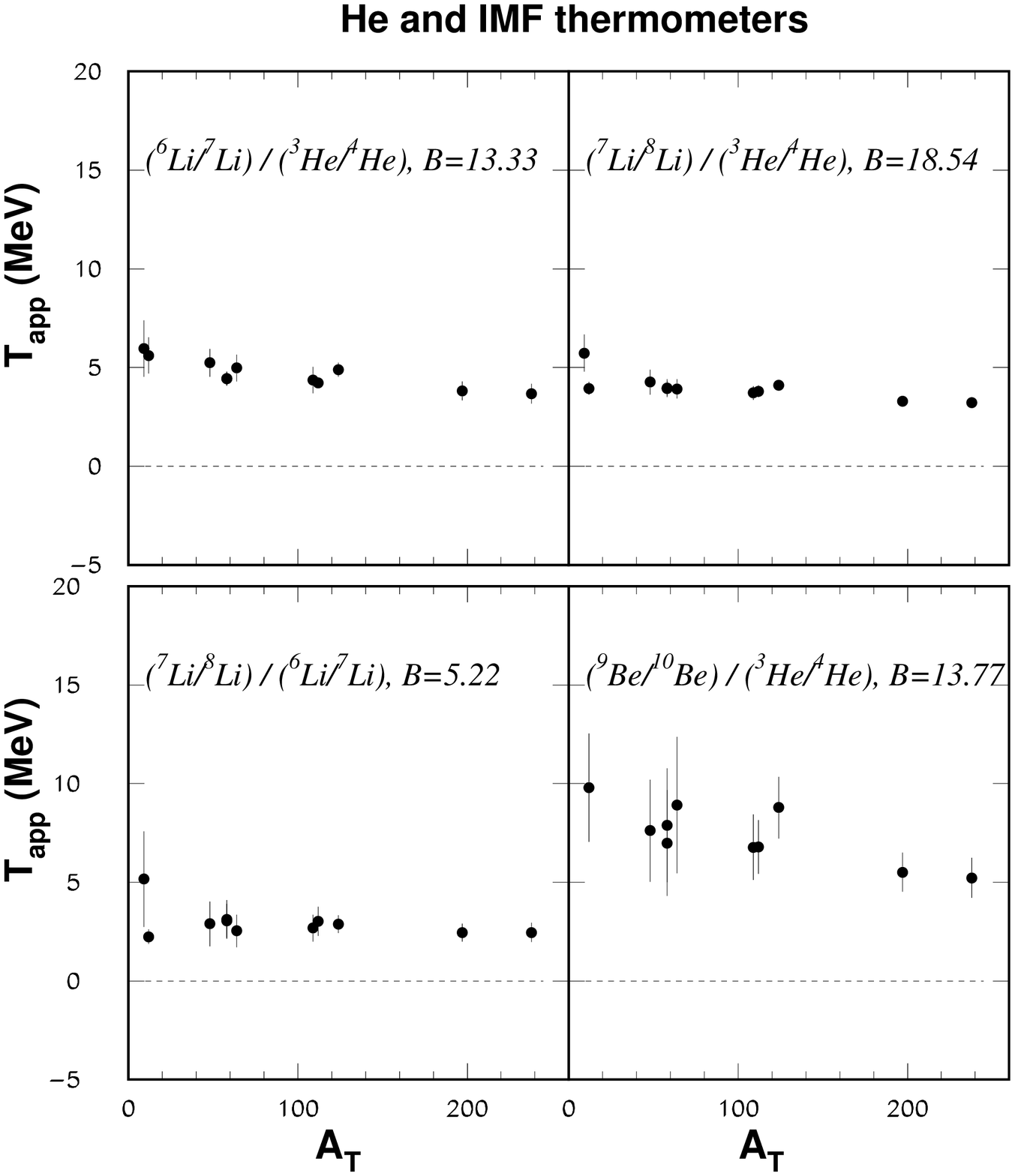,height=8.3cm}
\noindent
\small
\\
\begin{minipage}{7cm}
\vspace{-2mm}
\small {\bf Fig.~10.} Apparent temperatures obtained from hydrogen isotopes
as function of the target mass number A$_T$. Data from \cite{amn}
\end{minipage}
\normalsize
&
\begin{minipage}{7cm}
\small {\bf Fig.~11.} Apparent temperatures obtained from He and IMF isotopes 
as function of the A$_T$. Data from \cite{volnin,aln}
\end{minipage}
\normalsize
\end{tabular}
\end{center}

In the next step we converted the above values of $T_{app}$ into 
temperatures $T$ at the breakup point as far as the correction factors 
for sequential decay \cite{hxi} were available.
The mean correction amounts to $\simeq$ 5\% but does not exceed 15\%.

The target mass dependence for thermometers including isotopic pairs
with $^3$He/$^4$He is demonstrated in the Fig.~12.
Some features of Fig.~12 are worth to discuss.\\
{\bf (i)}~~ The temperatures which have been derived from the differential
cross sections at $\Theta_{lab}$=60$^\circ$ (open circles)
are larger in comparison with those obtained from production cross sections.
Enhanced temperatures in forward direction and strong variations of such 
thermometers which involve $^3$He/$^4$He ratios 
were also reported for heavy-ion induced reactions.\\
{\bf (ii)}~~ We observed some structures in the temperatures 
obtained from fragment yields at $\Theta_{lab}$=60$^\circ$ for $^{48}$Ti, 
$^{58}$Ni, $^{64}$Ni, $^{112}$Sn and $^{124}$Sn targets (open circles). 
These data~\cite{volnin} are characterized by the low cut-off
in the measured kinetic energy distributions and small corrections for the
missing part of the spectra to obtain energy integrated yields. 
Regardless of the indicated errors, Fig.~12 shows within a 
limited region of A$_T$ a common systematic trend to higher 
temperatures with increasing ratio $N_t/Z_t$ of the target. 
The isotopically separated targets $^{58,64}$Ni and
$^{112,124}$Sn form such groups which look like local fluctuations 
but more probably they are caused by 
different ${\cal N}$/${\cal Z}$ ratios of the fragmenting nuclei.\\
{\bf (iii)}~~ If we consider the full range of A$_T$ 
from Be to U, the opposite tendency dominates: 
the breakup isotopic temperature decreases weakly with increasing A$_T$. 
The temperatures which include the ratio $^3$He/$^4$He show the most pronounced
decrease with increasing A$_T$. This behavior seems to be related to
the exceptional properties of $^4$He which were attributed to a predominant
production by evaporation~\cite{volnin}.
\vspace{-10mm}   
\begin{center}
\begin{tabular}{cc}
\epsfig{file=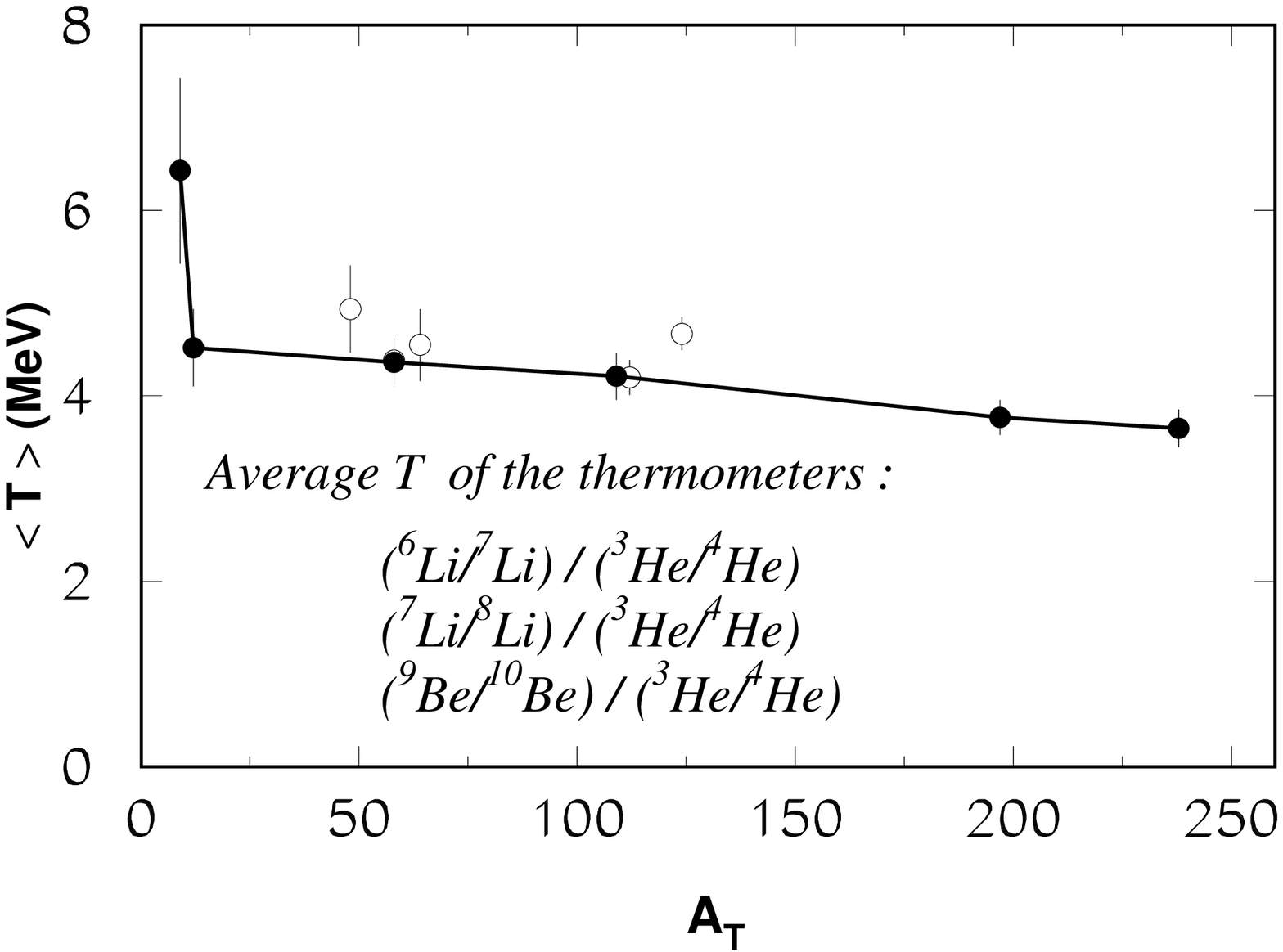,width=8cm}
\noindent
\small
&
\epsfig{file=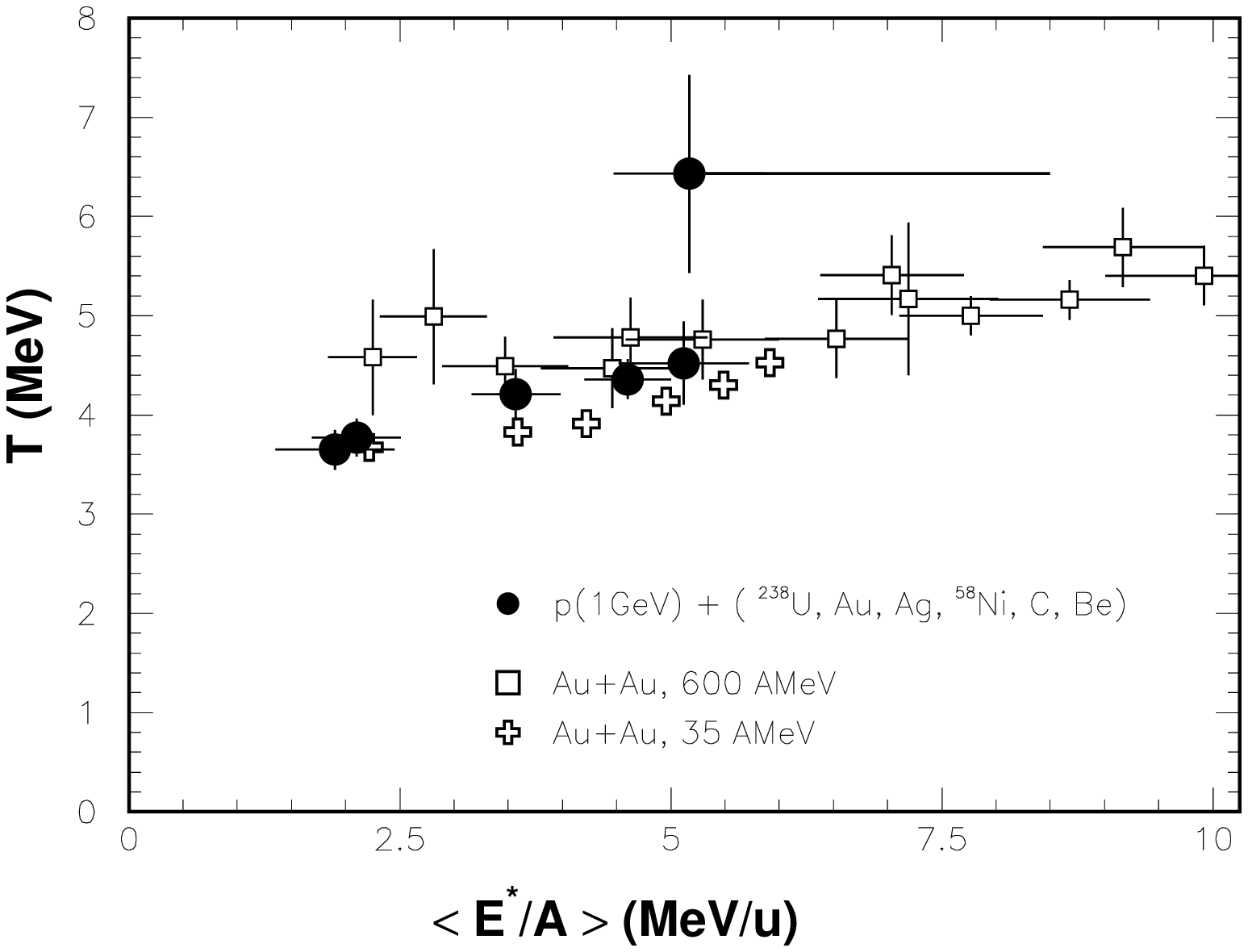,height=6.3cm} 
\noindent
\small
\\
\begin{minipage}{8cm}
\small {\bf Fig.~12.} Isotopic temperatures $T$ as function of target mass number.
         Open circles: data from measurements at $\Theta_{lab}$=60$^\circ$.
         Black dots connected by solid lines are evaluated from isotope production
         cross sections
\end{minipage}
\normalsize
&
\begin{minipage}{8cm}
\small {\bf Fig.~13.} Comparison of temperatures $T$ as function of the 
                      exciation energy E$^*$/A derived from 1-GeV proton 
                      induced fragmentation (black dots) with 
                      IMF production in Au+Au collisions presented in
                      the caloric curve \cite{poch,mil}.
\end{minipage}
\normalsize
\end{tabular}
\end{center}
On the other side, double ratios involving heavier isotopes (~Li,~Be,~B)
provide temperatures which are nearly independent 
on the target mass within the error bars  \cite{epja}.

Within the considered range of A$_T$ the average isotopic 
temperature $\langle T \rangle \simeq $4 MeV was obtained from 
measured cross sections (Fig.~12, black dots).
It should be noted here that thermometry analysis performed
for isotopic yields of ternary charged particles
accompanying low energy fission of heavy nuclei resulted in
significant lower temperature $\simeq $1 MeV \cite{epja2}.

Figure 13 gives another presentation of the obtained results:
the isotopic temperature against the excitation energy. 
The excitation energy $\langle E^*/A \rangle$ was estimated 
from earlier measurements of the mass loss and the linear momentum 
transfer for nuclei emitting IMF's \cite{kotov}.
The data obtained from 1 GeV proton induced fragmentation show 
an agreement with those derived from heavy ion collisions.
We mention the difference between the caloric curve,
which characterizes the temperature evolution of one system
as a function of the excitation energy and the black points 
which are presented in Fig.~13.
The latter one's belong to different excited systems 
in a wide range of target masses A$_T$.

\underline{Summarizing}, 
different isotope thermometers employing inclusive data obtained 
in \mbox{1 GeV} proton induced fragmentation,
involving various target nuclei, were analysed.

$\bullet$
It was found that even thermometers which involve isotope pairs with
{\em B} $<$ 10 MeV provide steady results which may be suitable 
for relative temperature measurements. 

$\bullet$
The estimated average breakup temperature  
$\langle T \rangle \simeq$ 4 MeV, derived from double isotopic 
yield ratios of the fragmentation products at 1 GeV proton energy,
is consistent with the corresponding one characterizing the plateau of 
the caloric curve, obtained for heavy-ion induced multifragmentation.

$\bullet$ The weak dependence of the isotopic temperatures
          on the target mass A$_T$ suggests speculations about
          a comprehensive applicability of thermodynamical concepts
          in nearly all nuclei involved in this analysis.

\section*{Overall conclusion}
\hspace*{6mm}
          Fragmentation and spallation reactions induced by 1 GeV protons
          show features of equilibration. The statistical mode 
          of fragment formation seems to prevail not only 
          in heavy ion reactions below $\simeq$100 $A\cdot$MeV 
          but also in proton-nucleus collisions in the GeV domain. 
Isospin degree of freedom plays an important role
in the description of residual nucleus and fragment production.

\end{document}